\documentclass[%
reprint,
superscriptaddress,
showkeys,
amsmath,
amssymb,
aps,
]{revtex4-1}

\usepackage{graphicx}
\usepackage{dcolumn}
\usepackage{bm}

\begin{document}

\preprint{APS/123-QED}

\title{Determination of the Equivalence Principle violation signal for the MICROSCOPE space mission: optimization of the signal processing}

\author{Emilie Hardy}%
\email{emilie.hardy@onera.fr}
\affiliation{ONERA, 29 avenue de la Division Leclerc, F-92322 Ch\^atillon, France}

\author{Agn\`es Levy}
\affiliation{ONERA, 29 avenue de la Division Leclerc, F-92322 Ch\^atillon, France}
\author{Gilles M\'etris}%
\affiliation{G\'eoazur, Universit\'e de Nice Sophia-Antipolis, Centre National de la Recherche Scientifique (UMR 7329), Observatoire de la C\^ote d'Azur, 250 avenue Albert Einstein, 06560 Valbonne, France}
\author{Manuel Rodrigues}%
\affiliation{ONERA, 29 avenue de la Division Leclerc, F-92322 Ch\^atillon, France}
\author{Pierre Touboul}
\affiliation{ONERA, 29 avenue de la Division Leclerc, F-92322 Ch\^atillon, France}

\date{September 24, 2013}

\begin{abstract}
The MICROSCOPE space mission aims at testing the Equivalence Principle (EP) with an accuracy of $10^{-15}$. The test is based on the precise measurement delivered by a differential electrostatic accelerometer on-board a drag-free microsatellite which includes two cylindrical test masses submitted to the same gravitational field and made of different materials. The experiment consists in testing the equality of the electrostatic acceleration applied to the masses to maintain them relatively motionless at a well-known frequency.
This high precision experiment is compatible with only very little perturbations. However, aliasing arises from the finite time span of the measurement, and is amplified by measurement losses. These effects perturb the measurement analysis. Numerical simulations have been run to estimate the contribution of a perturbation at any frequency on the EP violation frequency and to test its compatibility with the mission specifications. Moreover, different data analysis procedures have been considered to select the one minimizing these effects taking into account the uncertainty about the frequencies of the implicated signals.

\end{abstract}

\pacs{Valid PACS appear here}
\maketitle


\section{Introduction}
\label{intro}

The Equivalence Principle (EP) is at the basis of General Relativity and its main consequence is the Universality of Free Fall, that is to say that the acceleration of an object in a gravitational field is independent of its mass and its internal composition. It leads to the equivalence between the inertial mass (which measures the resistance of an object to acceleration) and the gravitational mass (which is used to compute the gravitational force exerted by or applied on an object). The Universality of Free Fall has been tested throughout the centuries with an improving accuracy. Lately, experiments using sophisticated torsion-balances have led to a record accuracy of a few $10^{-13}$ (\cite{Schlamminger-TorsionBalance}). However, the accuracy of these on-ground experiments is limited by the numerous perturbations of the terrestrial environment. It is necessary to test the Equivalence Principle with an even better accuracy because it is a direct test of Einstein General Relativity (\cite{Will-relativity}) and because some unification theories which try to merge gravitation with the three other fundamental interactions expect a violation of the EP below $10^{-14}$ (\cite{Damour-ViolationEquivalencePrinciple}). Being performed in space, the MICROSCOPE mission will be able to overcome these limitations in order to test the Equivalence Principle with an accuracy of $10^{-15}$ never reached before (\cite{Touboul-EPspace}).

MICROSCOPE is a 300 kg microsatellite developed by CNES to orbit around the Earth for a two years mission. The launch of the satellite is scheduled for 2016. The onboard payload is composed of two differential electrostatic accelerometers developed by ONERA, each one being composed of two imbricated cylindrical test masses in electrostatic levitation. The masses are surrounded by a set of electrodes. Their positions along the three axes are detected thanks to capacitive sensors and control loops with electrostatic actuation keep them concentric at the center of the accelerometer cage, so that they both follow the same trajectory. The same electrodes allow both the action and the detection of the mass position thanks to a frequency separation: the detection is performed with a $100 \; \mbox{kHz}$ pumping signal while the servo-loop channels exhibit frequency of a few Hertz. The generated voltage is proportional to the mass acceleration. The acceleration measurement is then subsampled to $4 \; \mbox{Hz}$. 
The two masses are concentric and therefore submitted to the same gravitational field. For one of the differential accelerometer, the two masses are made of different composition. A difference between the measured forces applied to maintain them on the same trajectory would therefore indicate a violation of the Universality of Free Fall, and thus of the EP. The second accelerometer includes two test masses with the same composition and enables to test the measurement accuracy (\cite{Touboul-microscope2012}). 

The potential EP violation signal is expected at a well identified frequency, $f_{EP}$. A special effort has been made to reduce all the perturbations at $f_{EP}$ (\cite{Touboul-microscope2009}). In particular, an in-flight calibration of the instrument will be performed in order to correct the scientific measurement from the perturbations due to the instrumental parameters which limit the measurement accuracy at $f_{EP}$ (\cite{Josselin-microscope2010}). However, two monochromatic signals at distinct frequencies are not fully uncorrelated over a finite duration, in contrary to what happens over an infinite duration. The consequence for the MICROSCOPE experiment is that a perturbation at a frequency different from the EP frequency will have a non-null contribution at $f_{EP}$. This contribution, which we call "projection", is proportional to the amplitude of the perturbation and depends on its frequency. It is therefore necessary to determine this proportionality factor in order to define the specifications on the perturbations amplitude at every frequency. Moreover, the study of this phenomenon demonstrates the possibility to adjust the observation time span with respect to the main perturbations period in order to reduce the projection rates for the most important perturbations. Finally, the projection is modified in case of missing data. This effect is studied in order to determine appropriate procedures to process the data in the presence of measurement losses which enable to keep low amplitudes of projection on $f_{EP}$.

This paper first presents the influence of the finite duration of the observation window on the determination of the EP violation parameter and then focuses on the method applied to reduce the impact of the main perturbations. It will then expose the influence of the measurement losses and the procedures selected to deal with them.

\section{Influence of the observation window}
\label{sec:1}

\subsection{Impact of a perturbation signal on the extraction of the EP violation parameter}
\label{subsec:1.1}

The signal provided by the MICROSCOPE instrument is the half difference of the measurement of the electrostatic acceleration (deduced from the measured voltage and instrument parameters) applied to the two test masses of a differential accelerometer to keep them relatively motionless. This differential acceleration includes the potential violation signal due to a non-null EP violation parameter $\delta = \frac{m_{g2}}{m_{I2}} - \frac{m_{g1}}{m_{I1}}$, where $m_g$ is the gravitational mass and $m_I$ is the inertial mass respectively of the internal test mass (index 1) and external test mass (index 2). A second order model of the measurement has been defined \citep{Levy-SF2A}. The cylindrical test masses can move and are electrostatically controlled along the six degrees of freedom (three translations and three rotations). For the EP test, the measurement along the $\vec{X}$ axis of the instrument (the axis of cylinders, whose sensitivity exhibits the best sensitivity) is used. The form of the acceleration measurement is:
\begin{equation}
\Gamma_{mes,dx} = \frac{1}{2} \cdot g_x \cdot \delta + \Gamma_{inst,x}
\end{equation}
with:
\begin{itemize}
   \item $g_x$ the component of the gravitational acceleration along the $\vec{X}$ axis of the instrument ;
   \item $\Gamma_{inst}$ the perturbative acceleration that reduces the accuracy of the EP parameter determination. This acceleration is composed of a stochastic part (noise of the instrument and of the drag-free system) that can be reduced by integrating the measurement over a large duration, and an harmonic part that is due to instrumental parameters (scale factors, off-centering of the test masses, misalignement and coupling between the axes, quadratical parameters). This last part is largely reduced by calibrating these parameters in order to correct the measurement from their contributions \citep{Hardy-calib}.
\end{itemize}

The objective of the MICROSCOPE mission is to detect a potential EP violation signal. The EP violation parameter $\delta$ is extracted from the observed measurement $\Gamma_{mes,dx,corr}$, called $S_{obs}$. The model of the measurement used for the analysis is, for all sample times $t_i$:
\begin{equation}
S_{obs}(t_i) = \delta S_{EP}(t_i)
\end{equation}
with:
\begin{itemize}
   \item $S_{obs} = \Gamma_{mes,dx,corr}$ the half difference of the measured acceleration after correction from the impacts of the instrumental parameters thanks to the in-orbit calibration of the instrument ;
   \item $S_{EP} = \frac{1}{2} \cdot g_x$ the half gravitational acceleration along the measurement axis. This acceleration is determined with a high precision by using a Earth gravity field model.
\end{itemize}

The Least Square estimated solution is:
\begin{equation}
\hat{\delta} = \frac{\langle S_{EP}, S_{obs} \rangle}{\langle S_{EP}, S_{EP} \rangle }
\end{equation}
with $\langle X, Y \rangle$ the scalar product of $X$ and $Y$ and N the length of the vectors :
$$\langle X, Y \rangle = \sum_{i=1}^{N} X(t_i)Y(t_i) $$
A disturbing signal $S_d$ which is not identified and not taken into account in the measurement model modifies the estimation by:
\begin{equation}
\Delta \delta = \frac{\langle S_{EP}, S_{d} \rangle}{\langle S_{EP}, S_{EP} \rangle }
\end{equation}

When we consider the simple but sufficient hypothesis of two masses falling along a circular orbit around the Earth, considered as a gravitational monopole, the EP signal $S_{EP}$ is a sine signal whose frequency $f_{EP}$ and phases $\phi_{EP}$ are well-known. Two different pointing of the satellite will be experimented : inertial and spinning. In the case of a satellite in inertial pointing, the gravitational field is modulated by the satellite orbital frequency. The test frequency $f_{EP}$ is therefore the orbital frequency ($1.6 \cdot 10^{-4} \; \mbox{Hz}$, corresponding to an orbital period of about $6000 \; \mbox{s}$). In the case of a spinning pointing, $f_{EP}$ is the sum of the orbital frequency and the spin frequency ($6,1 \cdot 10^{-4} \; \mbox{Hz}$, chosen so that $f_{EP}$ is close to the minimum of the instrumental noise).

The residual perturbative acceleration at $f_{EP}$ included in the observed signal $S_{obs}$ after calibration and correction of the measurement introduces an error on the determination of the EP violation parameter $\delta$. The performance of the satellite attitude and orbit control system, the performance of the instrument and its in-orbit calibration enable to keep this error compatible with the mission accuracy objective by keeping the residual perturbation at $f_{EP}$ to an acceptable level. However, perturbations of strong amplitude may appear at other frequencies and have an impact on the determination of the EP violation parameter.

For this paper, the disturbing signal $S_d$ is considered to be a monochromatic signal at a frequency different from $f_{EP}$.   
\begin{eqnarray}
S_{EP} & = & A_{EP} \sin(\omega_{EP} t + \phi_{EP}) \\
S_{d}  & = & A_{d} \sin(\omega_{d} t + \phi_{d})
\end{eqnarray}
The impact of the disturbing signal $\Delta \delta$ is proportional to $\frac{A_{EP}A_d}{A_{EP}^2} = \frac{A_d}{A_{EP}}$. For the sake of simplicity, we define the projection rate $\tau$ of a frequency $f_d$ on the test frequency $f_{EP}$ as the normalized component at $f_{EP}$ presented by a monochromatic perturbation at $f_d$ for a given analysis time span $T$. The projection rate is linked to the impact of a perturbation at $f_d$ on the estimation of the Equivalence Principle parameter at $f_{EP}$:
\begin{eqnarray}
\tau &=& \Delta \delta \cdot \frac{A_{EP}}{A_{d}} \nonumber \\
\tau &=& \frac{\langle \sin(\omega_{EP} t + \phi_{EP}), \sin(\omega_{d} t + \phi_{d}) \rangle}{\langle \sin(\omega_{EP} t + \phi_{EP}), \sin(\omega_{EP} t + \phi_{EP}) \rangle } = \frac{\langle s_{EP}, s_{d} \rangle}{\langle s_{EP}, s_{EP} \rangle } \nonumber \\
     & &
\end{eqnarray}
with $S_{EP} = A_{EP} s_{EP}$ and $S_{d} = A_{d} s_{d}$.

\subsection{Analytical approximation}
\label{subsec:1.2}

To correspond to the discrete measurement, the scalar product is discrete and defined on a finite number of frequencies. However, it is possible to approximate it with a continuous integral for an analytical evaluation.
\begin{equation}
\langle X,Y \rangle = \int_{t = 0}^{T} X(t) Y(t) dt
\end{equation}
with $T$ the analysis time span.

For a perturbation pulsation $\omega_d$ different from the EP pulsation $\omega_{EP}$:
\begin{eqnarray}
\langle s_{EP}, s_{d} \rangle & = & \int_{t = 0}^{T} \sin(\omega_{EP} t + \phi_{EP}) \sin(\omega_{d} t + \phi_{d}) dt \\ 
                              & = & \frac{1}{2}\frac{1}{\omega_{EP}-\omega_{d}}[\sin((\omega_{EP} - \omega_{d})t + (\phi_{EP} - \phi_{d}))]_0^T \nonumber \\
                              & - & \frac{1}{2}\frac{1}{\omega_{EP}+\omega_{d}}[\sin((\omega_{EP} + \omega_{d})t + (\phi_{EP} + \phi_{d}))]_0^T \nonumber \\
                              &   &
\end{eqnarray}
For the rest of this analysis, the measurement time span $T$ is supposed to be chosen very near an exact number of the EP period (see section \ref{sec:2} for the explanation of this choice):
\begin{equation}
\omega_{EP}(T - \Delta t_e) \leq 2k\pi \leq \omega_{EP}T
\end{equation}
with $\Delta t_e$ the measurement sampling period. $\Delta t_e = 0.25 \; \mbox{s}$ and $\Delta t_e \ll T$, we can therefore approximate : $\omega_{EP}T \approx 2k\pi$.

The result of the scalar product is then:\\
\begin{widetext}
\begin{eqnarray}
\langle s_{EP}, s_{d} \rangle & \approx & \frac{1}{2}\frac{1}{\omega_{EP}-\omega_{d}}[\sin(-\omega_{d}T+\phi_{EP}-\phi_{d}) - \sin(\phi_{EP}-\phi_{d})]  \nonumber \\ 
                              & - & \frac{1}{2}\frac{1}{\omega_{EP}+\omega_{d}}[\sin(\omega_{d}T+\phi_{EP}+\phi_{d}) - \sin(\phi_{EP}+\phi_{d})] 
\end{eqnarray}
\end{widetext}

Besides, the denominator of the projection rate is:
\begin{eqnarray}
\langle s_{EP}, s_{EP} \rangle & =       & \int_{t = 0}^{T} \sin^2(\omega_{EP} t + \phi_{EP}) dt \nonumber \\
                               & =       & \frac{1}{2}\left(T - \frac{1}{2\omega_{EP}}\sin(2\omega_{EP} + 2\phi_{EP}) \right) \nonumber \\
                               & \approx & \frac{T}{2}
\end{eqnarray}
because $T >> T_{EP} = \frac{2\pi}{\omega_{EP}}$.

The projection rate therefore is:
\begin{eqnarray}
\label{eq:tau}
\tau & \approx & \frac{1}{T}\frac{1}{\omega_{EP}-\omega_{d}}[\sin(-\omega_{d}T+\phi_{EP}-\phi_{d}) - \sin(\phi_{EP}-\phi_{d})]  \nonumber \\ 
     & - & \frac{1}{T}\frac{1}{\omega_{EP}+\omega_{d}}[\sin(\omega_{d}T+\phi_{EP}+\phi_{d}) - \sin(\phi_{EP}+\phi_{d})] \nonumber \\ 
     &   &
\end{eqnarray}
The error due to the approximation $\omega_{EP}T \approx 2k\pi$ is less than $\frac{\Delta t_e}{T}$, whose order of magnitude is $10^{-6}$ for short duration sessions and $10^{-7}$ for long duration sessions.  

The projection rate on the EP violation signal is inversely proportional to the measurement time span. It is null if the analysis time is an integer multiple of the disturbance period ($\omega_d T = 2k\pi, k \in \mathbb{Z}$). The two signals $s_{EP}$ and $s_d$ are indeed orthogonal if they both match with a discrete Fourier line over the observation window $T$. This property will be used in section \ref{sec:2} to optimize the choice of the analysis duration and of the satellite spin frequency in order to minimize the impact of the most important perturbations.

For perturbations at lower frequencies, $\omega_d$ is negligible compared to $\omega_{EP}$ and we can approximate:
\begin{eqnarray}
\tau_{LF} \approx \frac{1}{\omega_{EP}T}& [ \sin(-\omega_{d}T+\phi_{EP}-\phi_{d}) - \sin(\phi_{EP}-\phi_{d})  \nonumber \\
                                        & - \sin( \omega_{d}T+\phi_{EP}+\phi_{d}) - \sin(\phi_{EP}+\phi_{d})] \nonumber \\
                                        & 
\end{eqnarray}
\begin{equation}
\tau_{LF} \approx -\frac{4}{\omega_{EP}T} \cos\phi_{EP}\sin \left(\frac{\omega_{d}T}{2} \right) \cos \left(\frac{\omega_{d}T}{2}+\phi_d \right)
\end{equation}
The projection rate of the perturbation on $f_{EP}$ is dependent on the phases of both the EP violation signal and the perturbation signal. We are interested in the worst case projection rate, and it is therefore maximized as a function of $\phi_d$:
\begin{equation}
\tau_{LF,max} = \frac{4}{\omega_{EP}T}\left|\sin\left(\frac{\omega_dT}{2} \right)\right| |\cos\phi_{EP}|
\end{equation}
The projection rate oscillates as a function of the perturbation pulsation $\omega_d$, between null and maximal values, drawing an envelope curve which varies as the inverse of the measurement duration $T$:
\begin{equation}
\tau_{LF,env} = \frac{4}{\omega_{EP}T}|\cos\phi_{EP}|
\end{equation}
At lower frequencies, the envelope curve is therefore independent of the perturbation frequency.

For perturbation at higher frequencies, $\omega_{d} >> \omega_{EP}$ and the approximated projection rate is:
\begin{equation}
\tau_{HF} \approx -\frac{4}{\omega_{d}T} \sin\phi_{EP}\sin \left(\frac{\omega_{d}T}{2} \right) \sin \left(\frac{\omega_{d}T}{2}+\phi_d \right)
\end{equation}
The equation of the envelope curve (maximised as a function of $\phi_d$) is:
\begin{equation}
\tau_{HF,env} = \frac{4}{\omega_{d}T}|\sin\phi_{EP}|
\end{equation}
At higher frequencies, the slope of the envelope curve is therefore -1.

\subsection{Numerical simulation}
\label{subsec:1.3}

The Least Squares Method leads to the same solution when applying a discrete Fourier Transform to each side of the equation. The scalar product can therefore be computed in the Fourier domain:
\begin{eqnarray} 
\langle X,Y \rangle &=& \Re(FT(X)(f_{EP}))  \Re(FT(Y)( f_{EP})) \nonumber \\
                    &+& \Im(FT(X)( f_{EP})) \Im(FT(Y)( f_{EP}))
\end{eqnarray}
In the case of a sine signal $s = \sin(2\pi f t + \phi)$ and with a rectangular window ($W_{rect}(t_i) = 1 \mbox{ if } t_i \in [0,T], 0 \mbox{ otherwise}$):\\
\begin{widetext}
\begin{eqnarray}
FT(s)(f_{EP}) & = & \frac{1}{2i} \exp(-i\pi\frac{f_{EP}-f}{f_e}(N-1)) \exp(i\phi)\frac{\sin\left(\pi\frac{f_{EP}-f}{f_{e}}N \right)}{\sin\left(\pi\frac{f_{EP} - f}{f_e} \right)} \nonumber \\
 & - & \frac{1}{2i} \exp(-i\pi\frac{f_{EP}+f}{f_e}(N-1)) \exp(-i\phi)\frac{\sin\left(\pi\frac{f_{EP}+f}{f_{e}}N \right)}{\sin\left(\pi\frac{f_{EP} + f}{f_e} \right)}
\end{eqnarray}
\end{widetext}
with $N$ the number of samples and $f_e$ the sampling frequency.

This method is implemented in the numerical simulation in order to compute the projection rate for perturbations whose frequency ranges from $10^{-5} \; \mbox{Hz}$ to $0.1 \; \mbox{Hz}$, corresponding to the frequency range where the disturbing signals are expected. As shown by the analytical computation in section \ref{subsec:1.2}, the resulting projection rate $\tau$ is dependent on the phases of the two signals, $\phi_{EP}$ and $\phi_d$. The phase $\phi_{EP}$ of the potential EP violation signal will be known for the data analysis. Moreover, as the phase depends on the orientation of the satellite, we will be able to chose it; several criteria (not only the numerical aspects tackled on this paper) could lead to different choices, and in practice various phases will be tested.
In this study, a worst-case value has been used for the specifications definition, and the figures presented in this paper thus represent the maximal value $\tau_{max}$ obtained when $\phi_{EP}$ and $\phi_d$ vary between $0$ and $2\pi$.

\begin{figure*}
\includegraphics[width=0.7 \textwidth]{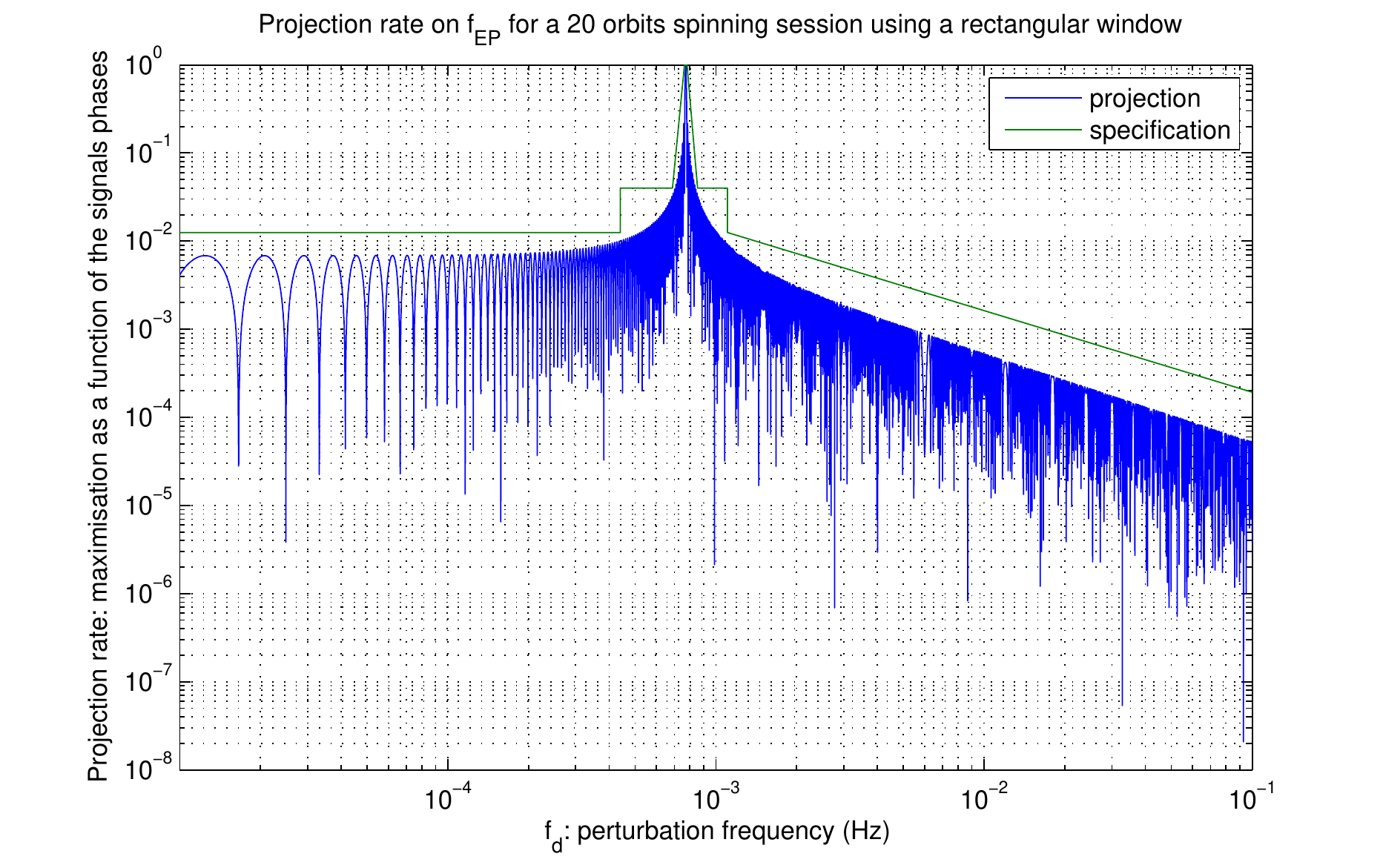}
\includegraphics[width=0.7 \textwidth]{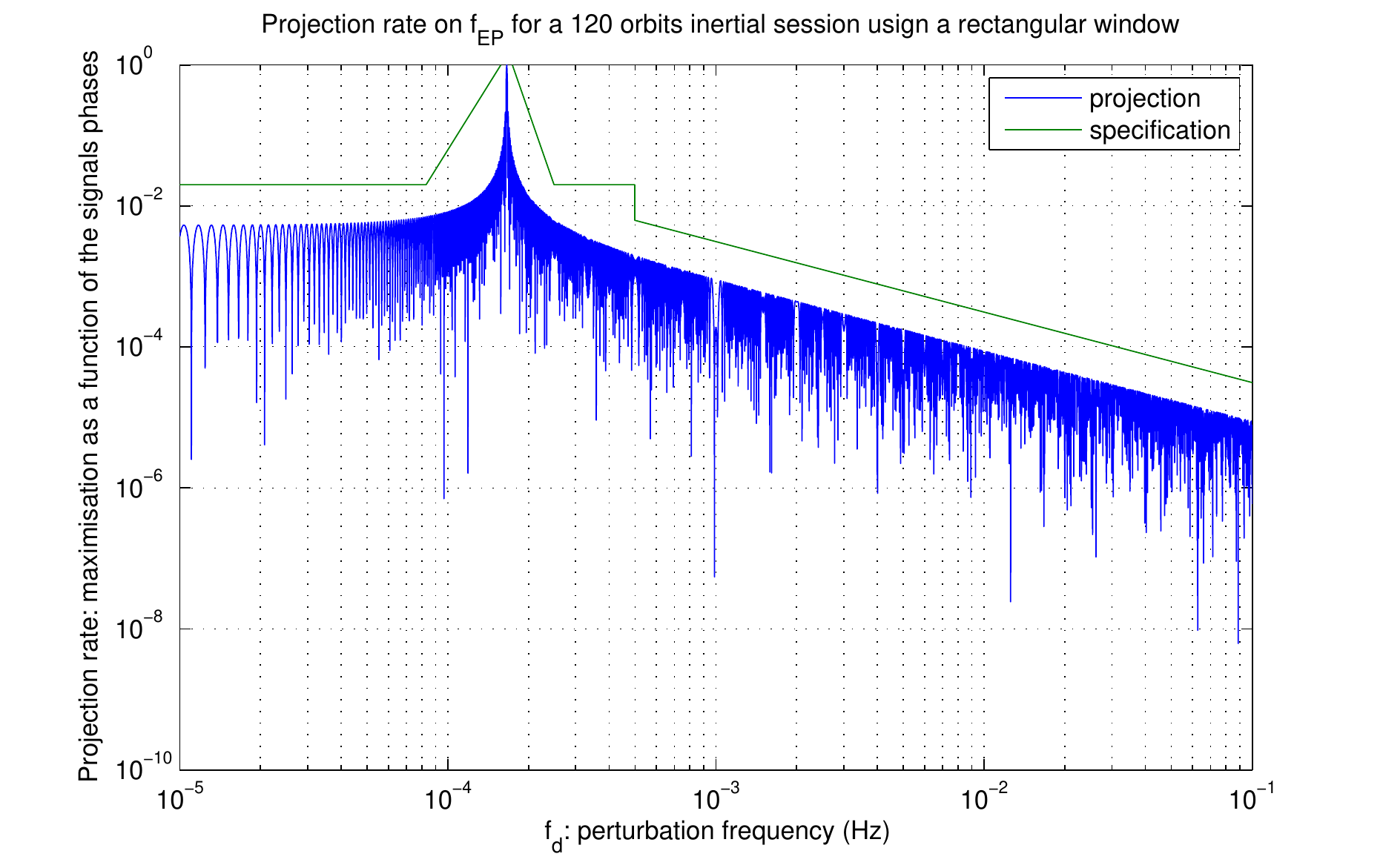}
\caption{\label{fig:gabarit_spinne} Projection rate of a perturbation on $f_{EP}$ as a function of the perturbation frequency in case of a rotating pointing of the satellite (a) or an inertial pointing (b). Note the 0 slope for low frequencies and the -1 slope for high frequencies.}
\end{figure*}

Figure \ref{fig:gabarit_spinne} presents the projection rate of a perturbation on $f_{EP}$ as a function of the perturbation frequency $f_d$ in case of a rotating pointing and an inertial pointing of the satellite. The variations of the curve correspond to those predicted by the analytical computation. The pattern has been defined from the envelope plot. It takes into account the local worst case value of the projection rate at each frequency in order to specify the amplitude of the perturbation signals that can be accepted for the experiment for a given duration of the observation window. Two different patterns have been defined for the rotating and the inertial pointing, because the implicated frequencies as well as the observation duration are different.

\subsection{Apodisation windows}
\label{subsec:1.4}

\begin{figure}
\centering
  \includegraphics[width=0.50\textwidth,clip]{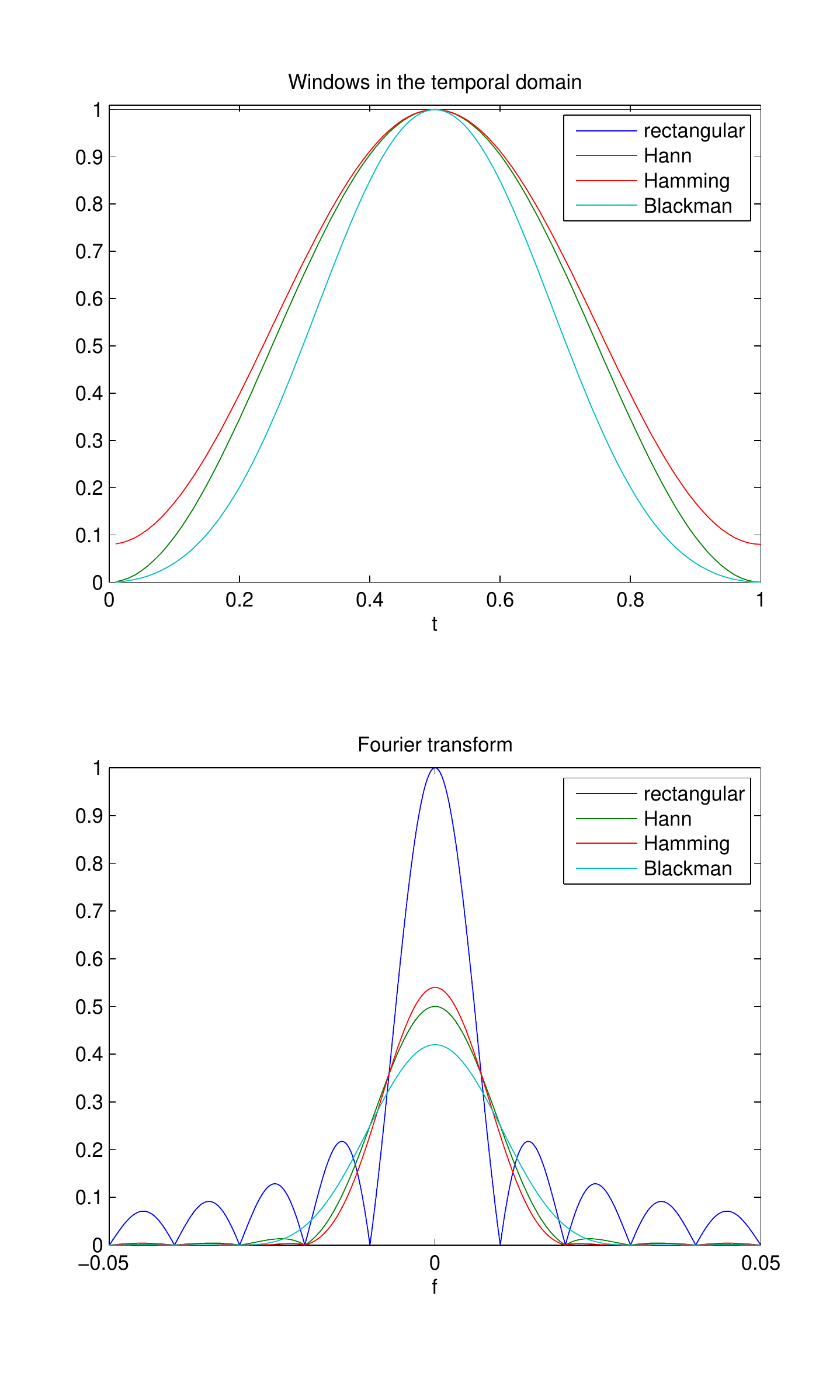}
\caption{Comparison between three apodisation windows and the rectangular window, in the temporal and Fourier domains.}
\label{fig:apodisation}       
\end{figure}

It is possible to use apodisation windows to considerably reduce the projection rate on $f_{EP}$. Three different windows have been tested for the extraction of the EP violation signal from the MICROSCOPE measurement: the Hann, the Hamming and the Blackman window, which are presented in figure \ref{fig:apodisation}. The principle of the apodisation is to make the signal converge slowly to zero at the edges of the window, in order to avoid the aliasing effects. As shown by figure \ref{fig:apodisation}, their effect in the Fourier domain is a much better attenuation of the side-lobes, and thus a reduction of the projection rates on the EP frequency.

For the first section of this paper, a simple rectangular window, $W_{rect}$, has been used. We will now study the effect of the Hann window:
\begin{equation}
W_{Hann} = \frac{1}{2} \left(1-\mbox{cos}\left(2\pi\frac{t}{T}\right) \right) = \frac{1}{2} \left(1 - \mbox{cos}(\omega_{mes}t) \right)
\end{equation}
where $T$ is the measurement time span and $\omega_{mes} = \frac{2\pi}{T}$. The same analytical computation as in section \ref{subsec:1.2} is developed to compute the temporal scalar product between the EP violation signal $S_{EP}$ and a perturbation signal $S_d$ with a Hann window:
\begin{equation}
\langle S_{EP},S_d \rangle = \int_{t = 0}^{T} S_{EP}(t) S_d(t) W_{Hann}(t) dt
\end{equation} 
The low frequency approximation gives:
\begin{equation}
\tau_{LF,Hann} = \tau_{LF,rect}\frac{1}{1-\left(\frac{\omega_{EP}}{\omega_{mes}} \right)^2} \approx \tau_{LF,rect} \left(\frac{\omega_{mes}}{\omega_{EP}} \right)^2
\end{equation} 
For the $120$ orbits inertial sessions, the global pattern of the projection rate on $f_{EP}$ is therefore $7 \times 10^{-5}$ times weaker when using a Hann window instead of a simple rectangular window.

At high frequencies, we can approximate:
\begin{equation}
\tau_{HF,Hann} = \tau_{HF,rect}\frac{1}{1-\left(\frac{\omega_{d}}{\omega_{mes}} \right)^2} 
\end{equation}
The ratio between the projection rates on $f_{EP}$ obtained with a rectangular window and those obtained with a Hann window increases as a function of the perturbation frequency.

\begin{figure*}
\includegraphics[width=0.7 \textwidth]{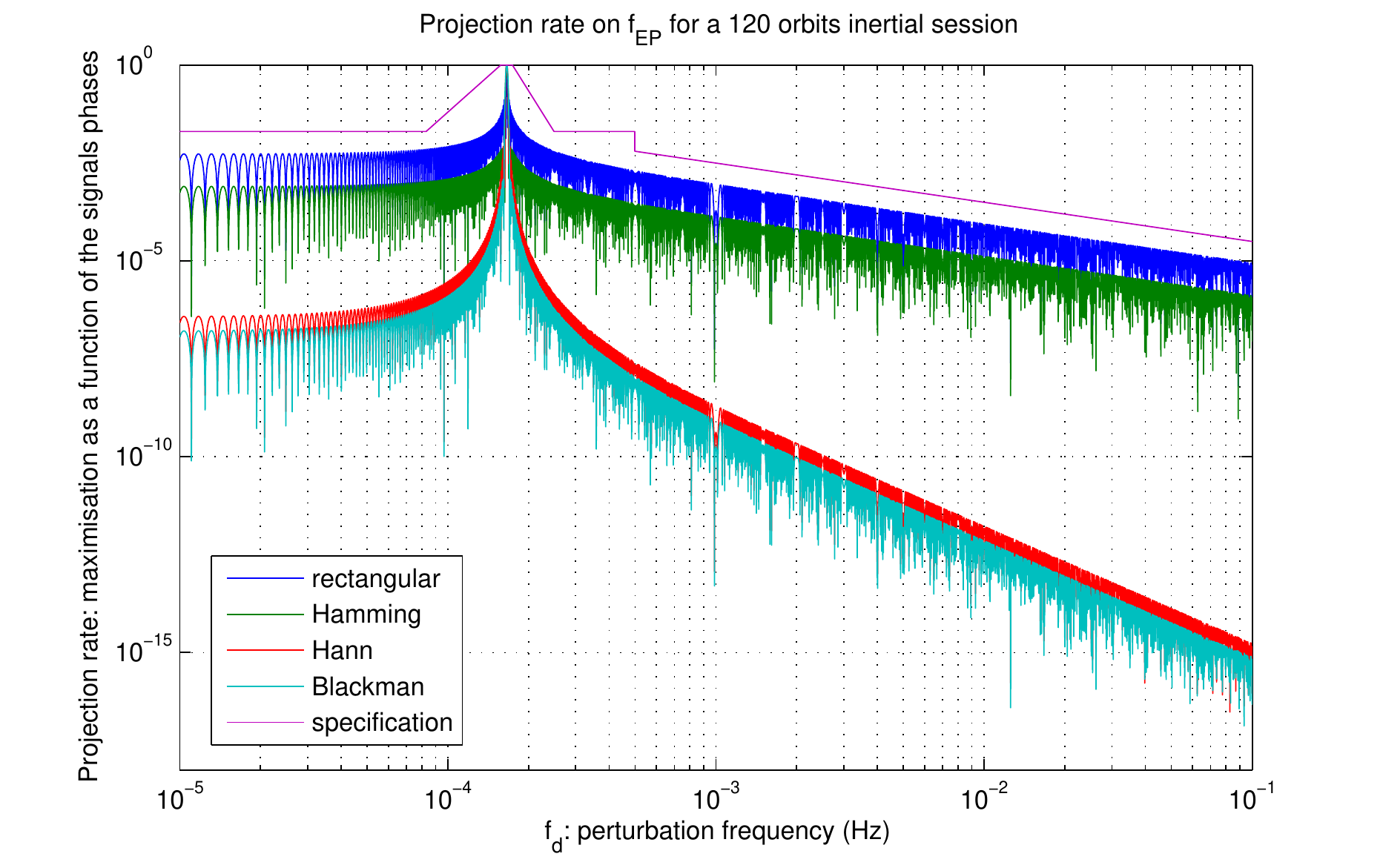}
\caption{\label{fig:apodisation-gabarit} Projection rate of a perturbation on $f_{EP}$ as a function of the perturbation frequency, using different windowing methods.}
\end{figure*}

Similar studies have been led for other apodisation windows. They have been confirmed by numerical simulations. As shown in figure \ref{fig:apodisation-gabarit}, the apodisation windows are effective in the case of the MICROSCOPE measurement. The best results are obtained with the Blackman window. However, we chose not to use any apodisation window for the simulations to follow. The definition of the specifications on the maximal amplitude of the perturbations is therefore based on the worst-case results of the projection rates. These specifications are used to define the satellite configuration. This choice avoids any a priori constraint on the future data process.

\section{Choice of the analysis time span and spin frequency}
\label{sec:2}

The most important perturbations (thermal effects, pointing variations...) occur at frequencies which are multiples of the orbital frequency in inertial pointing and linear combinations of the orbital and spin frequencies in rotating mode. These particular frequencies will be called singular frequencies in the rest of the paper:
\begin{equation}
 f_{d,sing} = n_1 f_{orb} + n_2 f_{spin}
 \end{equation}
with $n_1$ and $n_2$ being integers and $f_{spin}$ being possibly null in the case of an inertial pointing. These perturbations have such an amplitude that the direct application of the global pattern defined with the envelope curve of the projection rate (see figure \ref{fig:gabarit_spinne}) leads to projection amplitudes too high to enable us to reach the accuracy objective for the EP test.

The analysis of the performances at these singular frequencies leads to define more restrictive specifications on the projection rates than those of the global pattern. Some of these specifications are presented in table \ref{tab:freq_singulieres}, column 3.

\begin{table*}
\centering
\caption{Special specifications for some of the linear combinations of $f_{orb}$ and $f_{spin}$ in spinning pointing and projection rate in the worst case combination of errors on the determination of $f_{orb}$ and the realization of $f_{spin}$.}
\label{tab:freq_singulieres}       
\begin{tabular}{cccc}
\hline\noalign{\smallskip}
\hline
Frequency              & Global pattern value   & Specification         & Projection rate of    \\
                       &                        &                       & the singular frequencies \\
\noalign{\smallskip}\hline\noalign{\smallskip}
$f_{orb}$              & $1.25 \times{}10^{-2}$ & $10^{-4}            $ & $9.8 \times{}10^{-6}$ \\
$f_{spin} - 2 f_{orb}$ & $1.25 \times{}10^{-2}$ & $3.3 \times{}10^{-4}$ & $3.6 \times{}10^{-5}$ \\
$2 f_{orb}$            & $1.25 \times{}10^{-2}$ & $2.5 \times{}10^{-4}$ & $2.3 \times{}10^{-5}$ \\
$f_{spin} - f_{orb}$   & $4.0 \times{}10^{-2}$  & $3.3 \times{}10^{-4}$ & $3.2 \times{}10^{-5}$ \\
$3 f_{orb}$            & $4.0 \times{}10^{-2}$  & $5 \times{}10^{-4}  $ & $4.8 \times{}10^{-5}$ \\
$f_{spin}$             & $4.0 \times{}10^{-2}$  & $3.3 \times{}10^{-4}$ & $3.2 \times{}10^{-5}$ \\
$f_{spin} + f_{orb}$   & $1                  $  & $1                  $ & $1                  $ \\
$f_{spin} + 2 f_{orb}$ & $4.0 \times{}10^{-2}$  & $10^{-3}            $ & $7.9 \times{}10^{-5}$ \\
\hline\noalign{\smallskip}\hline
\end{tabular}
\end{table*}

The projection rates of these singular frequencies on $f_{EP}$ can be significantly reduced in order to comply with the specifications if we make sure that they match with discrete Fourier lines. In order to ensure this configuration for all those main perturbations, the analysis time span $T$ and the spin frequency have been chosen so that:
\begin{equation}
\label{eq:condition_rejection}
T = k_1 T_{orb} = k_2 T_{spin}
\end{equation}
with $k_1$ and $k_2$ being integers.
Equation \ref{eq:tau} demonstrates that this relation ensures a null value projection rate for all the perturbations at frequencies $f_{d,sing}$-like. In case of inertial pointing, the analysis time span has been chosen equal to $120$ orbits in order to reduce the instrumental stochastic noise to an acceptable level by integrating over this duration. In case of rotating pointing, the analysis time span can be reduced if the spin frequency is chosen so that the EP frequency is closer to the minimum of the instrumental noise which decreases as the inverse of the frequency. An appropriate compromise between the minimization of the instrumental noise and the technical constraints of the satellite attitude control system consists in choosing $k_2$ around $70$. It enables us to reduce the analysis duration to only 20 orbits.

\begin{figure}
\centering
  \includegraphics[width=0.5\textwidth,clip]{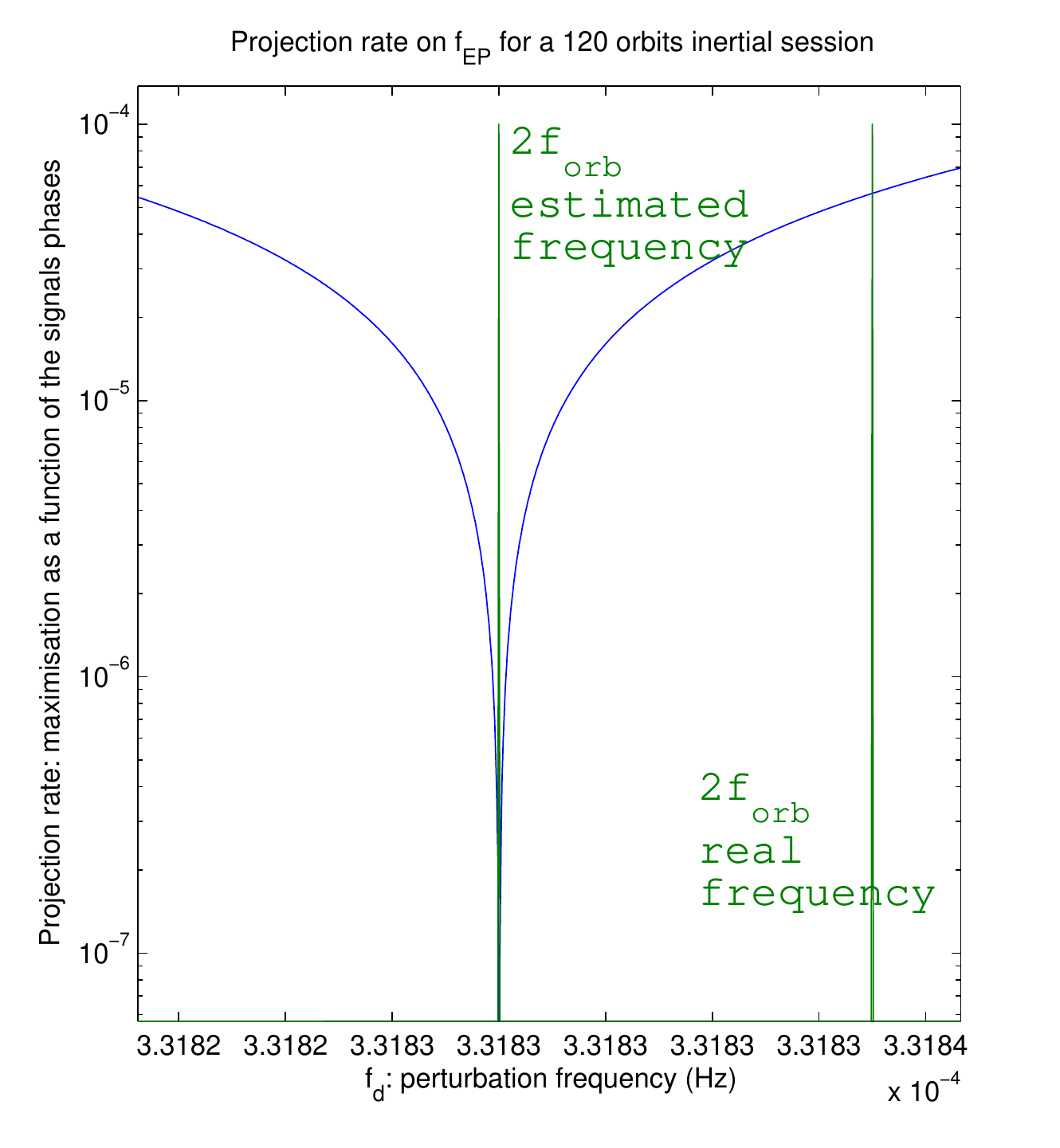}
\caption{Projection rate of a perturbation on $f_{EP}$ as a function of the perturbation frequency: zoom around $2f_{orb}$. The measurement time span $T$ has been optimized to correspond to an exact number of orbits. However, the estimated orbital frequency has been used to determine $T$. The real perturbation is at twice the real orbital frequency, and therefore does not match with the optimized projection rate.}
\label{fig:zoom_2forb}       
\end{figure}

However, the frequencies of the implicated signals are not perfectly known, leading to deviate slightly from the ideal case in which the projection rates are null for the singular frequencies. The uncertainty on the orbit determination causes an error on the orbital frequency which reaches $2 \times 10^{-8} \; \mbox{rad/s}$. The pointing of the satellite is realized by the attitude control system. The input of this system is the satellite attitude which is determined by using partly the measurement provided by the satellite star tracker and partly the angular acceleration measured by the differential accelerometer, which provides a better estimation of the attitude at high frequencies. Eight cold gas micro-thruster are then used to reach or keep the attitude instruction. The global specification for the accuracy of the attitude control system corresponds to a maximal error on the realization of the inertial pointing of $10^{-8} \; \mbox{rad/s}$ and on the command of the spinning pointing of $3 \times 10^{-8} \; \mbox{rad/s}$ (there are also specifications on the spin orientation, but they are not relevant for this paper). The effect of these uncertainties is illustrated in figure \ref{fig:zoom_2forb}. These spectral shifts have been introduced in the numerical simulation in order to get worst case projection rates. The results are gathered in table \ref{tab:freq_singulieres} and demonstrate the compatibility with the specifications.

\section{Measurement losses}
\label{sec:3}

The previous simulations have been run using the hypothesis of a regular sampling. In practice, measurement losses may appear. They will change the measurement sampling, and therefore the result of the scalar product of the perturbations and the EP signal, and therefore the projection rates.

\subsection{Origin of the measurement losses}
\label{subsec:3.1}

The measurement losses may be due to teletransmission errors. During the data transmission from the satellite to the ground station, part of the data may be lost. Part of the missing data could be recollected during the next fly-by, but it is impossible to guarantee a total recollection, leading to measurement losses whose duration ranges between a few seconds and a few hours. The experience of the PICARD mission provides a good estimation of the losses occurrence, since the MICROSCOPE satellite will follow nearly the same orbit and use the same station network.

Measurement may also become unexploitable in case of instrument saturation. It may happen because of crackles caused by decreasing pressure of the gas contained in the six thrust tanks of the satellite or because of the temperature changes of the satellite coating, which faces alternatively the Earth and the space vacuum. The resulting measurement losses are very short: shorter than one second.

Two procedures have been developed to deal with the measurement losses, depending on their duration.

\subsection{Short duration measurement losses}
\label{subsec:3.2}

For short measurement losses, two methods have been considered. The first one consists in simply replacing the missing data with null values. The appearance of measurement losses may either reduce or increase the projection rate on $f_{EP}$. But the projection rates for the main perturbations at the singular frequencies have been optimised to be as small as possible (see section \ref{sec:2}), and the measurement losses will therefore tend to increase them. The compatibility of the projection rates at the singular frequencies with the specifications is therefore the limiting factor that will constrain the maximal duration acceptable for a measurement loss.

In the case of telemetry losses, one measurement loss of variable duration is considered within the observation period. Numerical simulations have shown that the worst case corresponds to a measurement loss appearing in the middle of the measurement session. Table \ref{tab:rejection-trous} presents the projection rates for the main perturbations at the singular frequencies in the presence of a one minute long measurement loss in the middle of the 20 orbits session in rotating mode. These results are incompatible with the specifications. 

A possible correction consists in the reconstruction of the missing data by interpolating from the neighboring values. Actually a very simple interpolation is used: the mean value of the signal before and after the measurement loss is computed. Table \ref{tab:rejection-trous} shows that this second method is far more efficient than the first one. Different durations of the measurement loss have been tested: this method enables to accept one measurement loss whose duration is up to one minute per 20 orbits (up to one loss for the 20 orbits rotating sessions, six for the 120 orbits inertial sessions) in addition to the short coating and tanks crackles. This result has been obtained in the frame of a simulation without noise. In presence of the noise, the interpolation based only on one point before and after the interruption is probably not representative of the missing data. A solution to be considered will be to use the mean value of the signal over a duration long enough to reduce the noise to an acceptable level instead of only one point.

Moreover, a more sophisticated method is currently under study: the inpainting algorithm, which was first used in the image processing domain and has been successfully adapted to interpolate asteroseismic temporal missing data (\cite{Sato-Inpainting}). This method consists in representing the data in a dictionary where complete data are sparse and incomplete data are less sparse. 

\begin{table}
\centering
\caption{Projection rate and special specifications for some of the linear combinations of $f_{orb}$ and $f_{spin}$ in spinning pointing with a one minute measurement loss. Method 1: replacement of the lost data with null values; method 2: replacement of the lost data by the mean value of the signal before and after the measurement loss.}
\label{tab:rejection-trous}       
\begin{tabular}{cccc}
\hline\noalign{\smallskip}
                       & Projection            & Projection            &                       \\
Frequency              & rate:                 & rate:                 & Specification         \\
                       & method 1              & method 2              &                       \\
\noalign{\smallskip}\hline\noalign{\smallskip}
$f_{orb}$              & $7.4 \times{}10^{-4}$ & $9.8 \times{}10^{-6}$ & $10^{-4}            $ \\
$f_{spin} - 2 f_{orb}$ & $4.2 \times{}10^{-4}$ & $3.6 \times{}10^{-5}$ & $3.3 \times{}10^{-4}$ \\
$2 f_{orb}$            & $6.0 \times{}10^{-4}$ & $2.3 \times{}10^{-5}$ & $2.5 \times{}10^{-4}$ \\
$f_{spin} - f_{orb}$   & $3.8 \times{}10^{-4}$ & $3.2 \times{}10^{-5}$ & $3.3 \times{}10^{-4}$ \\
$3 f_{orb}$            & $1.8 \times{}10^{-4}$ & $4.8 \times{}10^{-5}$ & $5 \times{}10^{-4}  $ \\
$f_{spin}$             & $6.1 \times{}10^{-4}$ & $3.2 \times{}10^{-5}$ & $3.3 \times{}10^{-4}$ \\
$f_{spin} + f_{orb}$   & $1                  $ & $1                  $ & $1                  $ \\
$f_{spin} + 2 f_{orb}$ & $5.5 \times{}10^{-4}$ & $8.0 \times{}10^{-5}$ & $ 10^{-3}           $ \\
\hline\noalign{\smallskip}\hline
\end{tabular}
\end{table}

\subsection{Long duration measurement losses}
\label{subsec:3.3}

Because the rejection of the main perturbations at singular frequencies is the limiting factor, it is necessary to develop a procedure which enables to maintain low projection rates at these frequencies to be able to deal with measurement losses longer than one minute. To this end, the same method as the one described in section \ref{sec:2} can be used: the measurement time span and the adjustable frequencies are chosen so that the measurement duration corresponds to an entire number of the EP period and of the main perturbations periods. The strategy for long measurement losses consists in processing the data portions before and after the losses as independent sub-sessions, while making sure that each portion duration is adjusted to get the required orthogonality property between the EP signal and the main perturbations. In order to get the right analysis duration for the intermediate data portions, it may be necessary to expand the losses duration.

For the sessions in rotating pointing, this procedure is not applicable. According to section \ref{sec:2}, the measurement time span needs to correspond both to an entire number of the orbital period and of the spin period in order to correctly reject the main perturbations at the singular frequencies. 20 orbits being the least common multiple of these two periods, it is impossible to divide the session in shorter data portions that would still verify the required property. Fortunately, the probability of occurrence of a measurement loss whose duration is longer than one minute during a 20 orbit session is only $3 \%$. This is not negligible, but an unexploitable rotating session is far less disadvantaging than an unexploitable inertial session: they are short enough to be eventually run again.

\begin{figure*}[hbt]
\includegraphics[width=0.7 \textwidth]{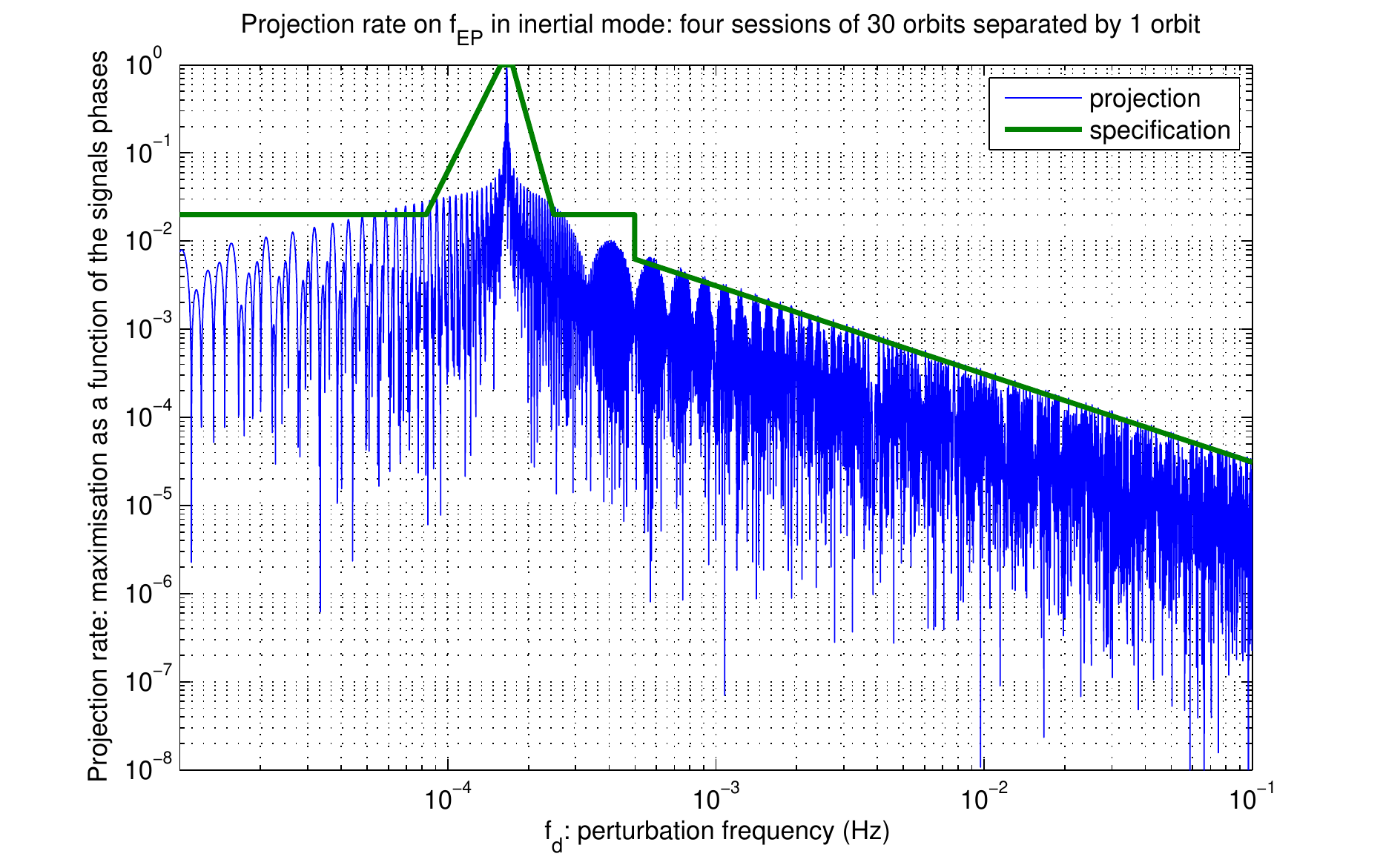}
\caption{\label{fig:gabarit-trouslongs} Projection rate of a perturbation on $f_{EP}$ as a function of the perturbation frequency: 120 orbits inertial session with three measurement losses whose duration is one orbit.}
\end{figure*}

In the case of the session in inertial pointing, the measurement time span is only required to correspond to an exact number of orbits in order to reduce the projection rates for the main perturbations at the singular frequencies. The strategy consists in eliminating from the measurement the entire orbit affected by the measurement loss, so that the data portions before and after the interruption correspond to an exact number of orbits, and therefore satisfy the property required for the singular frequencies. Numerical simulations have been run to test the number of long measurement losses that is acceptable: three measurement losses included in three suppressed periods of one orbit, as presented in figure \ref{fig:gabarit-trouslongs}. The probability of occurrence of more than three long measurement losses during a 120 orbits session is $0.4\%$, which is considered acceptable.

\section{Conclusion}

For the success of the MICROSCOPE space mission, a crucial problem is to limit the impact of possible perturbations by using appropriate methods of analysis. Because of the finite duration of the measurement window, a perturbation at any frequency may be projected at the well identified EP frequency. This phenomenon has been studied and a specification pattern has been defined in the two cases of the instrument attitude pointing with respect to the Earth gravity field: rotating pointing and inertial pointing. This pattern enables to take into account the projection rates of the perturbations all over the frequency spectrum. For the most important perturbations, additional more stringent specifications have been laid upon their projection rates.  It is therefore necessary to adjust the frequencies corresponding to the main perturbations, mainly the orbital and the spin frequency, to get a minimal projection. However, the projection effects are amplified by the frequencies uncertainties. Numerical simulations of the projection rate taking into account these uncertainties have proved the result to be compatible with the specifications. The projection can still be considerably improved by using apodisation windows, but the specifications are reached even in the worst case of a rectangular window.

The measurement losses increase the projection rate on the EP frequency. To deal with numerous very short losses (shorter than a second) or one loss up to a minute, the missing data are interpolated using the value of the measurement before and after the interruption. For measurement losses longer than a minute, the inertial session is separated in several independent data portions, each one verifying the required orthogonality property between the main perturbations frequencies and the EP frequency. Thanks to these procedures, the success probability of the mission reaches a level compatible with the specification.

These procedures will be implemented in the MICROSCOPE Scientific Mission Center (CMS) which is currently under development at ONERA.

\begin{acknowledgments}
The authors wish to thank the MICROSCOPE teams at CNES, OCA, ONERA and ZARM for the technical exchanges. This activity has received the financial support of Onera and CNES.
\end{acknowledgments}

\bibliography{hardy}
\bibliographystyle{unsrt}

\end{document}